\documentclass[10pt]{article}
\usepackage{graphicx}
\usepackage{slashed}
\usepackage{amsmath}

\begin{document}

\title{Operator Regularization of Feynman diagrams at multi-loop order}

\author{A.Y. Shiekh\footnote{\tt ashiekh@coloradomesa.edu} \\
              Department of Physics \\ 
              Colorado Mesa University \\
              Grand Junction, CO \\ 
              U.S.A.}

\date{}

\maketitle

\begin{abstract}
It may be possible to use operator regularization with Feynman diagrams, which would greatly simplify its use as it has so far been limited to the more complicated Schwinger approach. Operator regularization, unlike $\zeta$-function regularization, is not limited to one-loop order, and preserves supersymmetry unlike dimensional regularization. In practice the use of operator regularization in the context of Feynman diagrams is found not to complicate the calculation.
\end{abstract}

\section{Introduction}
Analytic regularization of quantum field-theory~\cite{Speer},~\cite{SalamStrathdee} is not new, but the operator regularization approach~\cite{McKeonSherry1},~\cite{McKeonSherry2} has not in general been used in conjunction with Feynman diagrams, although such use has been implied~\cite{Shiekh1},~\cite{Shiekh2}. Operator regularization has the distinct advantage that it can be used with formally non-renormalizable theories~\cite{Shiekh1},~\cite{MannEtAl} since the divergences are not reabsorbed, but each is removed and replaced by an arbitrary factor; so it might seem well worth the effort of having it work with Feynman diagrams, as it has so far been limited to the more complicated Schwinger approach.

While operator regularization does not cure the non-predictability problem of non-renormalizability, it has the advantage that the initial Lagrangian need not be extended, as would be the case for regularization approaches that do not remove the divergences and so would need the addition of extra terms in the original Lagrangian to accommodate them.

For use with Feynman diagrams, operator regularization in renormalizable theories needs to give results equivalent to other methods of regularization; the results will not be identical however, since operator regularization, unlike say dimensional regularization, removes divergences. The use of operator regularization with Feynman diagrams to one-loop was tackled previously~\cite{Shiekh3}.

We begin with a description of operator regularization, how it works, and why it should give equivalent results to dimensional regularization. This is followed by one and two-loop examples to show how operator regularization in practice is no harder then dimensional regularization.

The use of analytic continuation to deal with the divergences of quantum field theory has been criticized~\cite{Zee}, but even the formulae use in dimensional regularization to deal with all but the logarithmic divergence involve analytic continuation.

\section{Operator-Regularization}
The operator regularization scheme is governed by the identity:

\begin{equation}
\Omega^{-m} = 
\lim_{\varepsilon\rightarrow 0} \frac{d^n}{d\varepsilon^n}
\left(
( 1 + \alpha_1 \varepsilon + \cdots +\alpha_n \varepsilon^n )
\frac{\varepsilon^n}{n!} \Omega^{- \varepsilon -m}
\right)
\end{equation}
where the $\alpha_n$s are arbitrary, and it is enough that the degree of regularization ($n$) is the loop order.

There are two separate aspects to this procedure, first the regularization and then the continuation, where the divergences are replaced by arbitrary factors; these aspects could be separated if so desired.

\subsection{What operator regularization achieves}

Look at operator regularization for a divergent $\Omega^{-m}$

$$
\Omega^{-m} =
\lim_{\varepsilon \rightarrow 0}
\frac{d^n}{d \varepsilon^n}
\left(
(1+\alpha_1 \varepsilon+ \cdots +\alpha_n \varepsilon^n)\frac{\varepsilon^n}{n!} \Omega^{-\varepsilon-m}
\right)
$$
where the $\alpha_n$s are arbitrary.

Noting that $\Omega^{-\varepsilon-m}$ has the Laurent expansion

$$
\Omega^{-\varepsilon-m} =
\frac{1}{\varepsilon^n}c_{-n} + \cdots + \frac{1}{\varepsilon}c_{-1} + c_0 + \mathcal{O}(\varepsilon)
$$
one can now see that the effect of operator regularization is to replace the divergent poles by arbitrary constants

$$
\frac{1}{\varepsilon^n} \rightarrow \alpha_n
$$
to yield the finite interpretation

$$
\Omega^{-m} =
\alpha_n c_{-n} + \cdots + \alpha_1 c_{-1} + c_0
$$

\subsection{Generalization}
Operator regularization may be generalized to multiple operators, as appear in multi-loop cases.

\begin{equation}
A^{-k}\cdots Z^{-m} =
\lim_{\varepsilon \to 0} \frac{d^n}{d\varepsilon^n}
\left(
( 1 + \alpha_1 \varepsilon + \cdots +\alpha_n \varepsilon^n )
\frac{\varepsilon^n}{n!} A^{- \varepsilon -k}\cdots Z^{- \varepsilon -m}
\right)
\end{equation}
where the $\alpha_n$s are arbitrary, and it is sufficient that $n$ is the loop order;
this may be more compactly written as

$$
=
\left.
A^{- \varepsilon -k}\cdots Z^{- \varepsilon -m}
\right|_n
$$

\subsection{Feynman versus Schwinger}
The operator regularization method was first introduced in the context of the Schwinger approach, which while know to be equivalent to the Feynman approach, might still leave one asking if operator regularization has the same effect in both. So one is lead to asking if operator regularization of the logarithm as used in the Schwinger approach

\begin{equation}
\ln \Omega =
- \lim_{\varepsilon \rightarrow 0}
\frac{d^n}{d \varepsilon^n}
\left(
\frac{\varepsilon^{n-1}}{n!} \Omega^{-\varepsilon}
\right)
\label{eqn:Schwinger}
\end{equation}
is equivalent to operator regularization as used in the Feynman diagram context, namely:

\begin{equation}
\Omega^{-m} =
\lim_{\varepsilon \rightarrow 0}
\frac{d^n}{d \varepsilon^n}
\left(
\frac{\varepsilon^n}{n!} \Omega^{-\varepsilon-m}
\right)
\label{eqn:Feynman}
\end{equation}
The Schwinger form can be transformed into the Feynman form using

$$
\Omega^{-m} =
\frac{(-1)^{m-1}}{(m-1)!} \frac{d^m}{d \Omega^m} 
\ln \Omega
$$
to yield

$$
\Omega^{-m} =
\lim_{\varepsilon \rightarrow 0}
\frac{d^n}{d \varepsilon^n}
\left(
\frac{(1+\varepsilon)\cdots(m-1+\varepsilon)}{(m-1)!} 
\frac{\varepsilon^n}{n!}
\Omega^{-\varepsilon-m}
\right)
$$
which simplifies to

$$
\Omega^{-m} =
\lim_{\varepsilon \rightarrow 0}
\frac{d^n}{d \varepsilon^n}
\left(
\left(
1+\frac{\varepsilon}{1}
\right)
\cdots
\left(
1+\frac{\varepsilon}{m-1}
\right)
\frac{\varepsilon^n}{n!}
\Omega^{-\varepsilon-m}
\right)
$$
and can be seen to differ from the Feynman form (equation~\ref{eqn:Feynman}), so one might initially perceive a difference between the results of operator regularization in the Schwinger approach from that in the Feynman approach. However, so long as one includes all the arbitrary factors in operator regularization

\begin{equation}
\Omega^{-m} =
\lim_{\varepsilon \rightarrow 0}
\frac{d^n}{d \varepsilon^n}
\left(
(1+\overbrace{\alpha_1 \varepsilon+\cdots+\alpha_n \varepsilon^n}^\textit{\footnotesize resolutions of zero})
\frac{\varepsilon^n}{n!} \Omega^{-\varepsilon-m}
\right) 
\end{equation}
they then yield equivalent results.

\section{Equivalence at one-loop}

It is one thing that operator regularization should yield the same results for the Schwinger and Feynman approaches, but another that operator regularization yield equivalent results as other more well known regularization procedures such as dimensional regularization. This has been covered previously in the Schwinger approach~\cite{Rebhan} but the main purpose here is to use operator regularization with Feynman diagrams.

Starting from the basic integral that carries the divergences in one-loop Feynman diagrams

$$
\int \frac{d^{2\omega} l}{(2\pi)^{2\omega}} \frac{1}{\left(l^2+M^2+2l.p\right)^A} = 
\frac{\Gamma(A-\omega)}{\Gamma(A)} \frac{1}{\left(M^2-p^2\right)^{A-\omega}}
$$
the other versions follow by differentiating with respect to $p_\mu$, so one can concentrate on this one alone.

One needs to show that operator regularization and dimensional regularization treat the result

\begin{equation}
\frac{1}{(4\pi)^\omega} \frac{\Gamma(A-\omega)}{\Gamma(A)} \frac{1}{u^{A-\omega}}
\end{equation}
in equivalent ways when divergent, namely in the limit $A-\omega = 0, -1, \ldots$. 

In operator regularization this is

\begin{equation}
\frac{1}{(4\pi)^\frac{d}{2}} \lim_{\varepsilon \rightarrow 0} \frac{d}{d\varepsilon}
\left(
\frac{\varepsilon(1 + \alpha \varepsilon)}{\Gamma(A+\varepsilon)} \frac{\Gamma(\varepsilon-n)}{u^{\varepsilon-n}}
\right)
\end{equation}
while for dimensional regularization it is

\begin{equation}
\frac{1}{(4\pi)^{\frac{d}{2}-\varepsilon}}
\frac{\Gamma(\varepsilon-n)}{\Gamma(A)}
\frac{1}{u^{\varepsilon-n}}
\end{equation}
where $n \equiv \omega - A$ and having put $\omega=d/2$, where $d$ is the number of space-time dimensions. We are interested in comparing these when the original expression is divergent, namely when $n = 0, 1, \ldots$ etc.

The operator regularized expression evaluates in this case to (using the help of Mathematica~\cite{Wolfram})

\begin{equation}
\frac{(-1)^n}{(4\pi)^\frac{d}{2}} \frac{u^n}{(A-1)!n!}
\left(
\vphantom{\frac{1}{1}}
\alpha -\psi(A) + \psi(n+1) - \ln u
\right)
\end{equation}
while the dimensionally regularized expression has the Laurent expansion

\begin{equation}
\frac{(-1)^n}{(4\pi)^\frac{d}{2}} \frac{u^n}{(A-1)!n!}
\left(
\frac{1}{\varepsilon} + \ln 4\pi + \psi(n+1) - \ln u
\right) + \mathcal{O}(\varepsilon)
\end{equation}
where $\psi$ is the polygamma function.

These agree in form, and so all one-loop results should be equivalent when using operator regularization or dimensional regularization.

\subsection{Example: $\phi^3$ in 6 dimensions}
This investigation began with the suggestion\footnote{private communication with Professor D. G. C. McKeon} to look at the simpler divergent integral associated with scalar particles at one-loop in 6 dimensions, and continued with probing questions at the two-loop level.

Looking at this integral (rotated to Euclidean space)

\begin{equation}
\int \frac{d^6 l}{(2\pi)^6} \ 
\frac{1}{l^2+m^2} \frac{1}{(l+p)^2+m^2}
\end{equation}
This integral can be prepared by using the Feynman parameter `trick'

$$
\frac{1}{D_1^{a_1} D_k^{a_2}} = 
\frac{\Gamma(a_1+ a_2)}{\Gamma(a_1) \Gamma(a_2)}
\int_0^1 dx \ 
\frac{x^{a_1-1} (1-x)^{a_2-1}}{\left[D_1 x + D_2 (1-x)\right]^{a_1+ a_2}}
$$
to yield

\begin{equation}
= \int_0^1 dx \ 
\int \frac{d^6 l}{(2\pi)^6} \ 
\frac{1}{\left[l^2+m^2+p^2(1-x)+2l.p(1-x)\right]^2}
\label{eqn:starting}
\end{equation}
One can now proceed by comparing the use of operator regularization for this divergent integral to that of dimensional regularization.

\subsubsection{Operator regularization}
Using the generalized operator regularization scheme, the above becomes

$$
=
\int_0^1 dx \ 
\lim_{\varepsilon\rightarrow 0} \frac{d}{d\varepsilon}
\ \int \frac{d^6 l}{(2\pi)^6} \ 
\left( \frac{\varepsilon(1+\alpha \varepsilon)}{\left[l^2+m^2+p^2(1-x)+2 l.p(1-x)\right]^{\varepsilon+2}} \right)
$$
then performing the momentum integrals using the identity

\begin{equation}
\int \frac{d^{2\omega} l}{(2\pi)^{2\omega}} \ 
 \frac{1}{\left(l^2+M^2+2l . p\right)^A} = 
\frac{1}{(4\pi)^\omega \Gamma(A)} \frac{\Gamma(A-\omega)}{\left(M^2-p^2\right)^{A-\omega}}
\label{eqn:identity}
\end{equation}
leads to

$$
= \frac{1}{(4\pi)^3} \int_0^1 dx \ 
\lim_{\varepsilon\rightarrow 0} \frac{d}{d\varepsilon}
\left( 
\frac{\varepsilon(1+\alpha \varepsilon)}{\Gamma(\varepsilon+2)} 
\frac{\Gamma(\varepsilon-1)}{\left[m^2+p^2x(1-x)\right]^{\varepsilon-1}} 
\right)
$$
where one can now perform the operator regularization limit, using

$$
\lim_{\varepsilon\rightarrow 0} \frac{d}{d\varepsilon}
\left( 
\frac{\varepsilon(1+\alpha \varepsilon)}{\Gamma(\varepsilon+2)} 
\frac{\Gamma(\varepsilon-1)}{u^{\varepsilon-1}} 
\right)
=
u (-\alpha + \ln u)
$$
to yield the finite result

$$
= \frac{1}{(4\pi)^3}\int_0^1 dx \ 
\left(m^2+p^2 x(1-x)\right)
\left(-\alpha + \ln\left[m^2+p^2 x(1-x)\right] \right)
$$
This can be evaluated further, but this is not necessary for comparison with the result from dimensional regularization, but $\mu^2$ should be included in the logarithm, and taken from the arbitrary ($\alpha$) part, to yield the fixed part

\begin{equation}
\frac{1}{(4\pi)^3}\int_0^1 dx \ 
\left(m^2+p^2 x(1-x)\right)
\ln\left[\frac{m^2+p^2 x(1-x)}{\mu^2}\right]
\label{eqn:OpRegFixed}
\end{equation}
and the arbitrary part

\begin{equation}
- \frac{\alpha}{(4\pi)^3} \left( m^2 + \frac{p^2}{6} \right)
\label{eqn:OpRegArbitrary}
\end{equation}

\subsubsection{Dimensional regularization}
Regularization might now be done using the dimensional approach; starting from the same point (equation~\ref{eqn:starting}) with dimensional extension

$$
=
\int_0^1 dx \ 
\int \frac{d^{6-2\varepsilon}l}{(2\pi)^{6-2\varepsilon}} \ 
\frac{1}{\left[l^2+m^2+p^2(1-x)+2l.p(1-x)\right]^2}
$$
and again using the identity of equation~\ref{eqn:identity} leads to

$$
= \frac{\Gamma(-1+\varepsilon)}{(4\pi)^{3-\varepsilon}} 
\int_0^1 dx \ 
\frac{1}{\left[m^2+p^2x(1-x)\right]^{-1+\varepsilon}}
$$
Using Mathematica~\cite{Wolfram} to help expand about $\varepsilon = 0$

$$
\frac{\Gamma(-1+\varepsilon)}{(4\pi)^{3-\varepsilon}}
u^{1-\varepsilon}
= -\frac{1}{(4 \pi)^3} \left(\frac{1}{\varepsilon} - \gamma + 1 + \ln 4 \pi\right) u
+ \frac{u \ln u}{(4 \pi)^3} 
+ \mathcal{O}(\varepsilon)
$$
where $\gamma$ denotes the Euler-Mascheroni constant ($0.577\ldots$); things become, for the finite part (again including the $\mu^2$)

\begin{equation}
\frac{1}{(4\pi)^3}\int_0^1 dx \ 
\left( m^2 + p^2 x(1-x) \right)
\ln \left[\frac{m^2 + p^2 x(1-x)}{\mu^2}\right]
\end{equation}
leaving the divergent part

\begin{equation}
-\frac{1}{(4 \pi)^3} \left(\frac{1}{\varepsilon} - \gamma + 1 + \ln 4 \pi \right)
\left( m^2 + \frac{p^2}{6} \right)
\end{equation}
which are seen to agree in form with the result from operator regularization (equations~\ref{eqn:OpRegFixed} and~\ref{eqn:OpRegArbitrary}).

\subsection{Example: QED}
Applying operator regularization to the three divergent one-loop Feynman diagrams in QED, following Ramond~\cite{Ramond} with the Feynman gauge and in Euclidean space.

\subsubsection{One-loop correction to the fermion line}

Starting with the Feynman diagram for the one-loop correction to the fermion line ($\Sigma(p)$) (diagrams drawn using JaxoDraw~\cite{JaxoDraw})

\begin{center}
\includegraphics[scale=1]{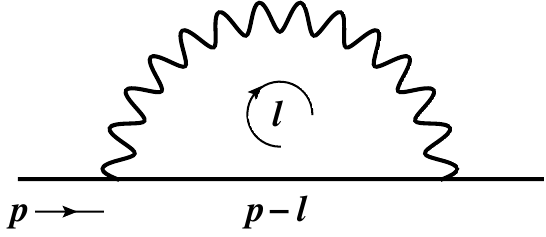}
\end{center}
\begin{equation}
\Sigma(p) = -e^2 \int \frac{d^4 l}{(2\pi)^4} \ 
\gamma_\mu \frac{-i}{\slashed{p}-\slashed{l}+m} \gamma_\nu 
\frac{\delta_{\mu\nu}}{l^2}
\end{equation}
Following Ramond~\cite{Ramond} this simplifies to

\begin{equation}
\Sigma(p) = - i e^2 \int^1_0 dx \ 
\int \frac{d^4 l}{(2\pi)^4} \ 
\frac{\gamma_\mu[\slashed{p}(1-x)-m]\gamma_\mu}{\left[l^2+m^2x+p^2x(1-x)\right]^2}
\label{eqn:fermionstarting}
\end{equation}
which is taken as the common starting point for both dimensional and operator regularization.

Now proceeding with operator regularization, following the same general route that would be taken with dimensional regularization

$$
\Sigma(p) = - i e^2 \int^1_0 dx \ 
\lim_{\varepsilon \rightarrow 0} \frac{d}{d\varepsilon}
\int \frac{d^4 l}{(2\pi)^4} \ 
\frac{\varepsilon(1+\alpha\varepsilon)\gamma_\mu[\slashed{p}(1-x)-m]\gamma_\mu}
{\left[l^2+m^2 x+p^2 x(1-x)\right]^{\varepsilon+2}}
$$
using the identity of equation~\ref{eqn:identity1} to perform the momentum integrals

$$
= - i \frac{e^2}{(4\pi)^2}
\int^1_0 dx \ 
\gamma_\mu[\slashed{p}(1-x)-m]\gamma_\mu
\lim_{\varepsilon \rightarrow 0} \frac{d}{d\varepsilon}
\left(
\frac{\varepsilon(1+\alpha\varepsilon)}{\Gamma(\varepsilon+2)} 
\frac{\Gamma(\varepsilon)}{\left[m^2 x+p^2 x(1-x)\right]^\varepsilon}
\right)
$$
Mathematica~\cite{Wolfram} for the limits

$$
\lim_{\varepsilon \rightarrow 0} \frac{d}{d\varepsilon}
\left(
\frac{\varepsilon(1+\alpha\varepsilon)}{\Gamma(\varepsilon+2)} 
\frac{\Gamma(\varepsilon)}{u^\varepsilon}
\right)
=
\alpha -1 - \ln u
$$
and the fact that in 4 dimensions $\gamma_\mu \gamma_\mu = -4$ 
and $\gamma_\mu \gamma_\rho \gamma_\mu = 2 \gamma_\rho$ (from $\{\gamma_\mu,\gamma_\nu\}=-2\delta_{\mu\nu}$) yields

$$
= - 2i \frac{e^2}{(4\pi)^2}
\int^1_0 dx \ 
\left(\slashed{p}(1-x)+2m\right)
\left(
\alpha-1-\ln\left[m^2x+p^2x(1-x)\right]
\right)
$$
to deliver the fixed part ($ \mu^2$ taken from the arbitrary $\alpha$)

\begin{equation}
2i \frac{e^2}{(4\pi)^2}
\int^1_0 dx \ 
\left(\slashed{p}(1-x)+2m\right)
\ln \left[
\frac{m^2x+p^2x(1-x)}{ \mu^2}
\right]
\end{equation}
and the arbitrary part

\begin{equation}
-i\frac{e^2}{(4\pi)^2} (\alpha-1) (\slashed{p}+4m)
\end{equation}
Compare this to the result from dimensional regularization~\cite{Ramond}, the finite part

\begin{equation}
2i \frac{e^2}{(4\pi)^2}
\int^1_0 dx \ 
\left(\slashed{p}(1-x)+2m\right)
\ln \left[
\frac{m^2x+p^2x(1-x)}{4\pi \mu^2}
\right]
\end{equation}
and the divergent part

\begin{equation}
-i \frac{e^2}{(4\pi)^2}
\left(
\frac{1}{\varepsilon} - \gamma - 1
\right)
(\slashed{p}+4m)
\underbrace{- i \frac{e^2}{(4\pi)^2} 2m}_\textit{extra term}
\label{eqn:extra}
\end{equation}
where $\gamma$ denotes the Euler-Mascheroni constant ($0.577\ldots$).

There is a constant difference between these two methods that stems from dimensionally continuing the gamma matrices (in dimensional regularization alone), but this will be absorbed by the counter terms.

\subsubsection{One-loop correction to the photon line}

Continuing with the diagram for the one-loop correction to the photon line ($\Pi_{\mu\nu}(p)$)

\begin{center}
\includegraphics[scale=1]{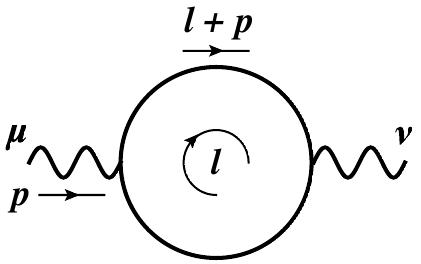}
\end{center}
\begin{equation}
\Pi_{\mu\nu}(p) = -e^2 \int \frac{d^4 l}{(2\pi)^4} \ {\rm Tr}
\left(
\gamma_\mu \frac{1}{\slashed{l}+\slashed{p}+m} \gamma_\nu \frac{1}{\slashed{l}+m}
\right)
\end{equation}
Following Ramond~\cite{Ramond} this simplifies to

\begin{equation}
\Pi_{\mu\nu}(p) = 8 e^2 (p_\mu p_\nu - \delta_{\mu\nu}p^2) \int_0^1 dx \ 
\int \frac{d^4 l}{(2\pi)^4} \ 
\frac{x(1-x)}{\left[l^2+m^2+p^2x(1-x)\right]^2}
\end{equation}
which is taken as the common starting point for both dimensional and operator regularization.

Proceeding with operator regularization, and again following the same general route that would be taken with dimensional regularization

$$
\Pi_{\mu\nu}(p) = 8 e^2 (p_\mu p_\nu - \delta_{\mu\nu}p^2) \int_0^1 dx \ 
\lim_{\varepsilon \rightarrow 0} \frac{d}{d\varepsilon}
\int \frac{d^4 l}{(2\pi)^4} \ 
\frac{\varepsilon(1+\alpha \varepsilon) \ x(1-x)}{\left[l^2+m^2+p^2x(1-x)\right]^{\varepsilon+2}}
$$
Performing the momentum integrals (using the identity of equation~\ref{eqn:identity1})

$$
= 8 \frac{e^2}{(4\pi)^2} (p_\mu p_\nu - \delta_{\mu\nu}p^2) \int_0^1 dx \ 
\lim_{\varepsilon \rightarrow 0} 
\frac{d}{d\varepsilon}
\left(
\frac{\varepsilon(1+\alpha\varepsilon)}{\Gamma(\varepsilon+2)}
\frac{\Gamma(\varepsilon)}{\left[m^2+p^2x(1-x)\right]^\varepsilon}
\right)
$$
again using

$$
\lim_{\varepsilon \rightarrow 0} \frac{d}{d\varepsilon}
\left(
\frac{\varepsilon(1+\alpha\varepsilon)}{\Gamma(\varepsilon+2)} \frac{\Gamma(\varepsilon)}{u^\varepsilon}
\right)
=
\alpha-1-\ln u
$$
yielding the finite part:

\begin{equation}
- 8 \frac{e^2}{(4\pi)^2} (p_\mu p_\nu - \delta_{\mu\nu}p^2) \int_0^1 dx \ 
\ln \left[ \frac{m^2+p^2x(1-x)}{\mu^2} \right]
\end{equation}
and the arbitrary part

\begin{equation}
\frac{4}{3} \frac{e^2}{(4\pi)^2} (p_\mu p_\nu - \delta_{\mu\nu}p^2) (\alpha-1)
\end{equation}
Compare this against the result of dimensional regularization; the finite part

\begin{equation}
- 8 \frac{e^2}{(4\pi)^2} (p_\mu p_\nu - \delta_{\mu\nu}p^2) \int_0^1 dx \ 
\ln \left[ \frac{m^2+p^2x(1-x)}{2\pi \mu^2} \right]
\end{equation}
and the divergent part

\begin{equation}
\frac{4}{3} \frac{e^2}{(4\pi)^2} (p_\mu p_\nu - \delta_{\mu\nu}p^2)
\left(
\frac{1}{\varepsilon} - \gamma
\right)
\end{equation}
which are seen to agree in form.

\subsubsection{One-loop correction to the vertex}

Lastly the Feynman diagram for the one-loop correction to the vertex ($\Gamma_\rho(p,q)$)

\begin{center}
\includegraphics[scale=1]{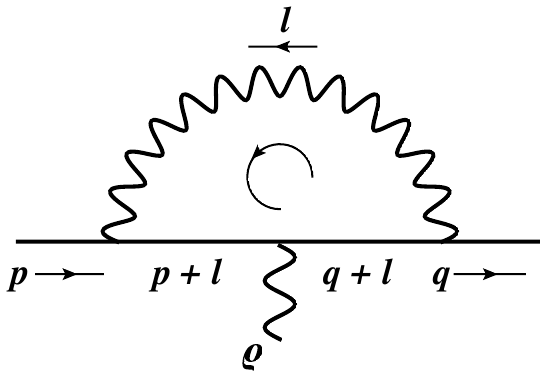}
\end{center}
\begin{equation}
\Gamma_\rho(p,q) = -i e^3 \int \frac{d^4 l}{(2\pi)^4} \ \gamma_\lambda
\frac{1}{\slashed{p}+\slashed{l}+m}
\gamma_\rho
\frac{1}{\slashed{q}+\slashed{l}+m}
\gamma_\sigma
\frac{\delta_{\lambda\sigma}}{l^2}
\end{equation}
Following Ramond~\cite{Ramond} and retaining only the divergent part for this investigation, this simplifies to

\begin{equation}
\Gamma^{(1)}_\rho(p,q) = -2i e^3 \int_0^1 dx \ \int_0^{1-x} dy \ 
\int \frac{d^4 l}{(2\pi)^4} \ 
\frac{\gamma_\sigma \slashed{l} \gamma_\rho \slashed{l} \gamma_\sigma}{\left(l^2+M^2\right)^3}
\end{equation}
where $M^2 \equiv m^2(x+y)+p^2x(1-x)+q^2y(1-y)-2p.q \ x y$ which is taken as the common starting point for both dimensional and operator regularization.

Proceeding with operator regularization, and again following the same general route that would be taken with dimensional regularization

$$
\Gamma^{(1)}_\rho(p,q) = -2i e^3 \int_0^1 dx \ \int_0^{1-x} dy \ 
\lim_{\varepsilon \rightarrow 0} \frac{d}{d\varepsilon} \int \frac{d^4 l}{(2\pi)^4} \ 
\varepsilon(1+\alpha \varepsilon)
\frac{\gamma_\sigma \slashed{l} \gamma_\rho \slashed{l} \gamma_\sigma}{\left(l^2+M^2\right)^{\varepsilon+3}}
$$
Performing the momentum integrals (using equation~\ref{eqn:identity3})

$$
= -i e \frac{e^2}{(4\pi)^2} \int_0^1 dx \ \int_0^{1-x} dy \ 
\lim_{\varepsilon \rightarrow 0} 
\frac{d}{d\varepsilon} 
\left(
\frac{\varepsilon(1+\alpha \varepsilon)}{\Gamma(\varepsilon+3)}
\frac{\Gamma(\varepsilon)}{\left(M^2\right)^\varepsilon}
\right)
\gamma_\sigma\gamma_\tau\gamma_\rho\gamma_\tau\gamma_\sigma
$$
then applying

$$
\lim_{\varepsilon \rightarrow 0} \frac{d}{d\varepsilon} 
\left(
\frac{\varepsilon(1+\alpha \varepsilon)}{\Gamma(\varepsilon+3)} \frac{\Gamma(\varepsilon)}{u^\varepsilon}
\right)
=
\frac{1}{4} (-3+2\alpha-2\ln u)
$$
with $\gamma_\sigma\gamma_\mu\gamma_\rho\gamma_\nu\gamma_\sigma = 2 \gamma_\nu\gamma_\rho\gamma_\mu$ and $\gamma_\mu\gamma_\rho\gamma_\mu = 2 \gamma_\rho$ leads to the finite part

\begin{equation}
2 i e \gamma_\rho \frac{e^2}{(4\pi)^2}
\int_0^1 dx \ \int_0^{1-x} dy \
\ln \left[ \frac{M^2}{ \mu^2} \right]
\end{equation}
where $M^2 \equiv m^2(x+y)+p^2x(1-x)+q^2y(1-y)-2p.q \ x y$,
and the arbitrary part

\begin{equation}
-i e \gamma_\rho \frac{e^2}{(4\pi)^2}
\left(
-\frac{3}{2}+\alpha
\right)
\end{equation}
Compare this to the result from dimensional regularization; the finite part

\begin{equation}
2 i e \gamma_\rho \frac{e^2}{(4\pi)^2}
\int_0^1 dx \ \int_0^{1-x} dy \
\ln \left[ \frac{M^2}{4\pi \mu^2} \right]
\end{equation}
where again $M^2 \equiv m^2(x+y)+p^2x(1-x)+q^2y(1-y)-2p.q \ x y$, and the divergent part

\begin{equation}
-i e \gamma_\rho \frac{e^2}{(4\pi)^2}
\left(
\frac{1}{\varepsilon} - \gamma -1
\right)
\end{equation}
which agree in form, recalling that operator regularization goes further than dimensional regularization in so much as that it actually removes the divergences.

\section{Equivalence at two-loop}
Not surprisingly, demonstrating equivalence at two-loop is somewhat more challenging than at one-loop.

\subsection{Example: Two-scoop diagram in $\phi^4$}
Starting with the two-scoop Feynman diagram for $\phi^4$

\begin{center}
\includegraphics[scale=1]{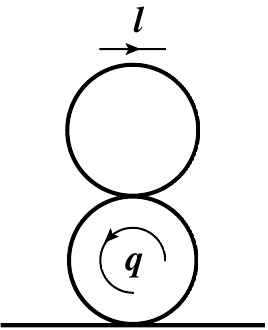}
\end{center}
In Euclidean-space this diagram

\begin{equation}
\frac{\lambda^2}{4}
\int \frac{d^4 l}{(2\pi)^4} \frac{1}{l^2+m^2}
\int \frac{d^4 q}{(2\pi)^4} \frac{1}{\left(q^2+m^2\right)^2}
\end{equation}
becomes under operator-regularization

$$
=
\frac{\lambda^2}{4}
\left.
\int \frac{d^4 l}{(2\pi)^4} \frac{1}{\left(l^2+m^2\right)^{\varepsilon+1}}
\int \frac{d^4 q}{(2\pi)^4} \frac{1}{\left(q^2+m^2\right)^{\varepsilon+2}}
\right|_2
$$
It is important to regulate the entire expression and not sub-parts 
(which would give a different result). Evaluating the integrals using the identity of equation~\ref{eqn:identity1} yields

$$
=
\frac{\lambda^2}{4}
\left.
\frac{\Gamma(\varepsilon-1)}{(4\pi)^2 \Gamma(\varepsilon+1)}\frac{1}{\left(m^2\right)^{\varepsilon-1}}
\frac{\Gamma(\varepsilon)}{(4\pi)^2 \Gamma(\varepsilon+2)}\frac{1}{\left(m^2\right)^{\varepsilon}}
\right|_2
$$
and finally performing the regularization (with the help of Mathematica~\cite{Wolfram}) to get the result

\begin{equation}
=
-\frac{\lambda^2 m^2}{1024 \pi^4}
\left(
1+\alpha_2 + 2 \alpha_1 \ln \frac{\mu^2}{m^2} + 2 \ln^2 \frac{\mu^2}{m^2}
\right)
\end{equation}
which agrees with the dimensional regularization result from Ramond~\cite{Ramond}, if

$$
\alpha_1 \to
\frac{1}{\varepsilon} + 1 - 2 \gamma
$$
and

$$
\alpha_2 \to
\frac{1}{\varepsilon^2} + \frac{1}{\varepsilon} ( 1 - 2\gamma ) -
2 \gamma (1-\gamma) + \frac{\pi^2}{6}
$$
where $\gamma$ denotes the Euler-Mascheroni constant ($0.577\ldots$). Where we have $\mu^2$, Ramond has $4\pi \mu^2$. So one sees that
the divergences have been correctly replaced by corresponding arbitrary factors.

\subsection{Example: Setting-Sun diagram in $\phi^4$}
Continuing with the first truly two-loop diagram

\begin{center}
\includegraphics[scale=1]{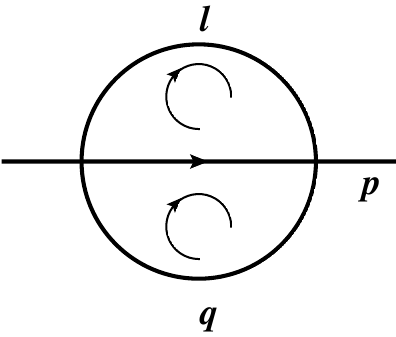}
\end{center}
Following Ramond~\cite{Ramond}

\begin{equation}
\Sigma(p) = - \frac{\lambda^2}{6}
\left(
3 m^2 K(p) + p^\mu K_\mu(p)
\right)
\end{equation}
where

\begin{equation}
K(p) = \int \frac{d^4 l}{(2\pi)^4} \int \frac{d^4 q}{(2\pi)^4}
\frac{1}{\left(q^2+m^2\right)^2(l^2+m^2)
[(q-l+p)^2+m^2]}
\end{equation}
and

\begin{equation}
K_\mu(p) = \int \frac{d^4 l}{(2\pi)^4} \int \frac{d^4 q}{(2\pi)^4}
\frac{(p+q-l)_\mu}{(q^2+m^2)(l^2+m^2)
\left[(q-l+p)^2+m^2\right]^2}
\end{equation}

\subsubsection{$K(p)$}
Again it is important to regulate the entire expression and not sub-parts 
(which would give a different result).

\begin{multline}
K(p) =
\\
\int \frac{d^4 l}{(2\pi)^4} 
\int \frac{d^4 q}{(2\pi)^4}
\left.
\frac{1}{\left[
\left(q^2+m^2\right)^2(l^2+m^2)((q-l+p)^2+m^2)
\right]^{\varepsilon+1}}
\right|_2
\end{multline}
Introducing Feynman parameters

\begin{multline*}
=
\int \frac{d^4 q}{(2\pi)^4}
\frac{1}{\left(q^2+m^2\right)^{2\varepsilon+2}}
\int \frac{d^4 l}{(2\pi)^4} 
\\
\left.
\frac{\Gamma(2\varepsilon+2)}{\Gamma^2(\varepsilon+1)}
\int_0^1 dx \ 
\frac{x^\varepsilon(1-x)^\varepsilon}{\left[(l^2+m^2)x+((q-l+p)^2+m^2)(1-x)\right]^{2\varepsilon+2}}
\right|_2
\end{multline*}
completing the square in $l$

\begin{multline*}
=
\frac{\Gamma(2\varepsilon+2)}{\Gamma^2(\varepsilon+1)}
\int_0^1 dx \ x^\varepsilon(1-x)^\varepsilon
\int \frac{d^4 q}{(2\pi)^4}
\frac{1}{\left(q^2+m^2\right)^{2\varepsilon+2}}
\\
\left.
\int \frac{d^4 l}{(2\pi)^4}
\frac{1}{\left[(l- (p+q)(1-x))^2+m^2+(p+q)^2 x(1-x)\right]^{2\varepsilon+2}}
\right|_2
\end{multline*}
changing variables $l' \equiv l - (p+q)(1-x)$ and performing the $l'$ integration (using the identity of equation~\ref{eqn:identity1})

\begin{multline*}
=
\frac{1}{(4 \pi)^2} \frac{\Gamma(2\varepsilon)}{\Gamma^2(\varepsilon+1)}
\int_0^1 dx \ x^\varepsilon(1-x)^\varepsilon
\\
\left.
\int \frac{d^4 q}{(2\pi)^4}
\frac{1}{\left(q^2+m^2\right)^{2\varepsilon+2}}
\frac{1}{\left[m^2+(p+q)^2x(1-x)\right]^{2\varepsilon}}
\right|_2
\end{multline*}
which can be rearranged as

\begin{multline*}
=
\frac{1}{(4 \pi)^2} \frac{\Gamma(2\varepsilon)}{\Gamma^2(\varepsilon+1)}
\int_0^1 dx \ \left(x(1-x)\right)^{-\varepsilon}
\\
\left.
\int \frac{d^4 q}{(2\pi)^4}
\frac{1}{\left[(p+q)^2+\frac{m^2}{x(1-x)}\right]^{2\varepsilon}}
\frac{1}{\left(q^2+m^2\right)^{2\varepsilon+2}}
\right|_2
\end{multline*}
Now introduce a second Feynman parameter

\begin{multline*}
=
\frac{1}{(4 \pi)^2}
\frac{\Gamma(4\varepsilon+2)}{\Gamma^2(\varepsilon+1) \Gamma(2\varepsilon+2)}
\int_0^1 dx\ \left(x(1-x)\right)^{-\varepsilon}
\int_0^1 dy\ y^{2\varepsilon-1}(1-y)^{2\varepsilon+1}
\\
\left.
\int \frac{d^4 q}{(2\pi)^4}
\left[
\left(
(p+q)^2 + \frac{m^2}{x(1-x)}
\right)y
+
(q^2+m^2)(1-y)
\right]^{-4\varepsilon-2}
\right|_2
\end{multline*}
completing the square in $q$

\begin{multline*}
=
\frac{1}{(4 \pi)^2}
\frac{\Gamma(4\varepsilon+2)}{\Gamma^2(\varepsilon+1) \Gamma(2\varepsilon+2)}
\int_0^1 dx\ \left(x(1-x)\right)^{-\varepsilon}
\int_0^1 dy\ y^{2\varepsilon-1}(1-y)^{2\varepsilon+1}
\\
\left.
\int \frac{d^4 q}{(2\pi)^4}
\left[
(q+p y)^2 + p^2 y(1-y) + m^2 \left(1-y+\frac{y}{x(1-x)} \right)
\right]^{-4\varepsilon-2}
\right|_2
\end{multline*}
changing variables $q' \equiv q+p y$ and performing the $q'$ integration (using the identity of equation~\ref{eqn:identity1})

\begin{multline*}
=
\frac{1}{(4 \pi)^4} 
\frac{\Gamma(4\varepsilon)}{\Gamma^2(\varepsilon+1) \Gamma(2\varepsilon+2)}
\int_0^1 dx\ \left(x(1-x)\right)^{-\varepsilon}
\int_0^1 dy\ y^{2\varepsilon-1}(1-y)^{2\varepsilon+1}
\\
\left.
\left[
p^2 y(1-y) + m^2\left(1-y+\frac{y}{x(1-x)} \right)
\right]^{-4\varepsilon}
\right|_2
\end{multline*}
The $y$ integral has a divergence at $\varepsilon =0$ which needs to be exposed, using

$$
y^{\varepsilon-1} = \frac{1}{\varepsilon} \frac{d}{dy} y^\varepsilon
$$
and integrating by parts, one finds

\begin{multline}
\overset{\rm op-reg}{K(p)} =
\\
-\frac{1}{(4 \pi)^4} 
\frac{\Gamma(4\varepsilon)}{\Gamma^2(\varepsilon+1) \Gamma(2\varepsilon+2)}
\frac{1}{2\varepsilon}
\int_0^1 dx\ \left(x(1-x)\right)^{-\varepsilon}
\int_0^1 dy\ y^{2\varepsilon}
\\
\left.
\frac{d}{dy}
\left(
(1-y)^{2\varepsilon+1}
\left[
p^2 y(1-y) + m^2\left(1-y+\frac{y}{x(1-x)} \right)
\right]^{-4\varepsilon}
\right)
\right|_2
\end{multline}
which can be compared to the result from dimensional regularization (Ramond~\cite{Ramond})

\begin{multline}
\overset{\rm dim-reg}{K(p)} =
\\
-\frac{\Gamma(2\varepsilon)}{(4 \pi)^{4-2\varepsilon}} 
\frac{1}{\varepsilon}
\int_0^1 dx\ \left(x(1-x)\right)^{-\varepsilon}
\int_0^1 dy\ y^{\varepsilon}
\\
\frac{d}{dy}
\left(
(1-y)
\left[
p^2 y(1-y) + m^2\left(1-y+\frac{y}{x(1-x)} \right)
\right]^{-2\varepsilon}
\right)
\end{multline}
The finite parts are now much more complicated than in the one-loop cases and it is prudent to proceed by showing that the difference between the results is zero. Using Mathematica~\cite{Wolfram} (code given in the appendix) to expand the expressions as a Laurent series in $\varepsilon$ and recalling that operator regularization replaces its $1/\varepsilon^2$ divergence with $\alpha_2$ and its $1/\varepsilon$ divergence with $\alpha_1$ one gets, for $(4\pi)^4$ times the difference:

\begin{multline*}
\left(
\frac{\alpha_1}{2}-\frac{1}{\varepsilon} - 2(1-\gamma+\ln 4\pi)
\right) \ln m^2 +
\\
\frac{\alpha_1}{4} - \frac{\alpha_2}{8} + \frac{1}{2\varepsilon^2}
+
\frac{1+ 2(\ln 4\pi-\gamma)}{2\varepsilon}
+\gamma^2+\frac{\pi^2}{12}+(1-2\gamma)\ln 4\pi+\ln^2 4\pi
\\
+ \mathcal{O}(\varepsilon)
\end{multline*}
One then gets a zero difference, and so agreement, if

$$
\alpha_1 \to \frac{2}{\varepsilon} + 4(1-\gamma+\ln 4\pi)
$$
and

\begin{multline*}
\alpha_2 \to 
\frac{4}{\epsilon^2} + 
\frac{8(1-\gamma+\ln 4\pi)}{\varepsilon}+
8\left(
1-2\gamma+\gamma^2+2(1-\gamma)\ln 4\pi + \ln^2 4 \pi
\right)+\frac{2}{3}\pi^2
\end{multline*}
where $\gamma$ denotes the Euler-Mascheroni constant ($0.577\ldots$). To agree with Ramond, we have also left out the $\mu^2$ factors. So one again sees that
the divergences have been correctly replaced by corresponding arbitrary factors.

\subsubsection{$K_\mu(p)$}
Starting with $K_\mu(p)$ and as before regulating

\begin{multline}
K_\mu(p) = 
\\
\int \frac{d^4 l}{(2\pi)^4} 
\int \frac{d^4 q}{(2\pi)^4}
\left.
\frac{(p+q-l)_\mu}{\left[(q^2+m^2)(l^2+m^2)\left((q-l+p)^2+m^2\right)^2\right]^{\varepsilon+1}}
\right|_2
\end{multline}
Introducing Feynman parameters

\begin{multline*}
= 
\int \frac{d^4 q}{(2\pi)^4}
\frac{1}{\left(q^2+m^2\right)^{\varepsilon+1}}
\int \frac{d^4 l}{(2\pi)^4} 
\\
\left.
\frac{\Gamma(3\varepsilon+3)}{\Gamma(2\varepsilon+2)\Gamma(\varepsilon+1)}
\int_0^1 dx \ 
\frac{x^\varepsilon(1-x)^{2\varepsilon+1}(p+q-l)_\mu}{\left[(l^2+m^2)x+((q-l+p)^2+m^2)(1-x)\right]^{3\varepsilon+3}}
\right|_2
\end{multline*}
completing the square in $l$

\begin{multline*}
= \frac{\Gamma(3\varepsilon+3)}{\Gamma(2\varepsilon+2)\Gamma(\varepsilon+1)}
\int_0^1 dx \ x^\varepsilon(1-x)^{2\varepsilon+1}
\int \frac{d^4 q}{(2\pi)^4}
\frac{1}{\left(q^2+m^2\right)^{\varepsilon+1}}
\\
\left.
\int \frac{d^4 l}{(2\pi)^4} 
\frac{(p+q-l)_\mu}{\left[(l-(p+q)(1-x))^2+m^2 +(p+q)^2 x(1-x)\right]^{3\varepsilon+3}}
\right|_2
\end{multline*}
letting $l' \equiv l - (p+q)(1-x)$

\begin{multline*}
= \frac{\Gamma(3\varepsilon+3)}{\Gamma(2\varepsilon+2)\Gamma(\varepsilon+1)}
\int_0^1 dx \ x^\varepsilon(1-x)^{2\varepsilon+1}
\int \frac{d^4 q}{(2\pi)^4}
\frac{1}{\left(q^2+m^2\right)^{\varepsilon+1}}
\\
\left.
\int \frac{d^4 l'}{(2\pi)^4}
\frac{(p+q)_\mu x - l'_\mu}{\left[{l'}^2+m^2 +(p+q)^2 x(1-x)\right]^{3\varepsilon+3}}
\right|_2
\end{multline*}
performing the $l'$ integral using the identity of equation~\ref{eqn:identity1} (dropping odd integrals)

\begin{multline*}
= \frac{1}{(4\pi)^2} 
\frac{\Gamma(3\varepsilon+1)}{\Gamma(2\varepsilon+2)\Gamma(\varepsilon+1)}
\int_0^1 dx \ x^{\varepsilon+1} (1-x)^{2\varepsilon+1}
\int \frac{d^4 q}{(2\pi)^4}
\frac{1}{\left(q^2+m^2\right)^{\varepsilon+1}}
\\
\left.
\frac{(p+q)_\mu}{\left[m^2 +(p+q)^2 x(1-x)\right]^{3\varepsilon+1}}
\right|_2
\end{multline*}
rearranging

\begin{multline*}
= \frac{1}{(4\pi)^2} 
\frac{\Gamma(3\varepsilon+1)}{\Gamma(2\varepsilon+2)\Gamma(\varepsilon+1)}
\int_0^1 dx \ x^{-2\varepsilon} (1-x)^{-\varepsilon}
\\
\left.
\int \frac{d^4 q}{(2\pi)^4}
\frac{(p+q)_\mu}{\left[(p+q)^2 + \frac{m^2}{x(1-x)}\right]^{3\varepsilon+1} \left(q^2+m^2\right)^{\varepsilon+1}}
\right|_2
\end{multline*}
Now introduce a second Feynman parameter

\begin{multline*}
= \frac{1}{(4\pi)^2} 
\frac{\Gamma(4\varepsilon+2)}{\Gamma(2\varepsilon+2)\Gamma^2(\varepsilon+1)}
\int_0^1 dx \ x^{-2\varepsilon} (1-x)^{-\varepsilon}
\int_0^1 dy \ y^{3\varepsilon} (1-y)^{\varepsilon}
\\
\left.
\int \frac{d^4 q}{(2\pi)^4}
\frac{(p+q)_\mu}{\left[ \left((p+q)^2 + \frac{m^2}{x(1-x)}\right)y + (q^2+m^2)(1-y)\right]^{4\varepsilon+2}}
\right|_2
\end{multline*}
completing the square in $q$

\begin{multline*}
= \frac{1}{(4\pi)^2} 
\frac{\Gamma(4\varepsilon+2)}{\Gamma(2\varepsilon+2)\Gamma^2(\varepsilon+1)}
\int_0^1 dx \ x^{-2\varepsilon} (1-x)^{-\varepsilon}
\int_0^1 dy \ y^{3\varepsilon} (1-y)^{\varepsilon}
\\
\left.
\int \frac{d^4 q}{(2\pi)^4}
\frac{(p+q)_\mu}{\left[ (q+p y)^2 + m^2\left( 1-y+\frac{y}{x(1-x)}\right) + p^2 y(1-y)\right]^{4\varepsilon+2}}
\right|_2
\end{multline*}
letting $q' \equiv q + p y$

\begin{multline*}
= \frac{1}{(4\pi)^2} 
\frac{\Gamma(4\varepsilon+2)}{\Gamma(2\varepsilon+2)\Gamma^2(\varepsilon+1)}
\int_0^1 dx \ x^{-2\varepsilon} (1-x)^{-\varepsilon}
\int_0^1 dy \ y^{3\varepsilon} (1-y)^{\varepsilon}
\\
\left.
\int \frac{d^4 q'}{(2\pi)^4}
\frac{p_\mu (1-y) + q'_\mu}{\left[ {q'}^2 + m^2\left( 1-y+\frac{y}{x(1-x)}\right) + p^2 y(1-y)\right]^{4\varepsilon+2}}
\right|_2
\end{multline*}
performing the $q'$ integral using the identity of equation~\ref{eqn:identity1} (dropping odd integrals), one finds

\begin{multline}
\overset{\rm op-reg}{K_\mu(p)} =
\\
p_\mu 
\frac{1}{(4\pi)^4} 
\frac{\Gamma(4\varepsilon)}{\Gamma(2\varepsilon+2)\Gamma^2(\varepsilon+1)}
\int_0^1 dx \ x^{-2\varepsilon} (1-x)^{-\varepsilon}
\int_0^1 dy \ y^{3\varepsilon} (1-y)^{\varepsilon+1}
\\
\left.
\left[p^2 y(1-y) + m^2\left( 1-y+\frac{y}{x(1-x)}\right) \right]^{-4\varepsilon}
\right|_2
\end{multline}
which can be compared to the result from dimensional regularization (Ramond~\cite{Ramond})

\begin{multline}
\overset{\rm dim-reg}{K_\mu(p)} =
\\
p_\mu 
\frac{\Gamma(2\varepsilon)}{(4\pi)^{4-2\epsilon}} 
\int_0^1 dx \ (x(1-x))^{-\varepsilon}
\int_0^1 dy \ y^{\varepsilon} (1-y)
\\
\left[p^2 y(1-y) + m^2\left( 1-y+\frac{y}{x(1-x)}\right) \right]^{-2\varepsilon}
\end{multline}
Proceeding as in the $K(p)$ case, by looking at the difference, the two regularizations agree if

$$
\alpha_1 \to \frac{2}{\varepsilon} + 5 + 4(2\ln 2+\ln \pi - \gamma)
$$
where $\gamma$ denotes the Euler-Mascheroni constant ($0.577\ldots$). So one again sees that
the divergences have been correctly replaced by corresponding arbitrary factors.

\section{Conclusion}
The above suggests that operator regularization can in fact be used in conjunction with Feynman diagrams to all loop orders. The calculation using operator regularization is actually somewhat simpler than that using dimensional regularization, as the gamma matrices are not dimensionally continued when using operator regularization.

While the main purpose of this work is to propose the possibility of using operator regularization in the context of Feynman diagrams, it has been noted that the results of operator regularization and dimensional regularization may differ~\cite{Rebhan}, and that dimensional regularization can have problems respecting supersymmetry~\cite{McKeonEtAl}.

\appendix
\section{Appendix}

\subsection{Feynman parameters}

\begin{equation}
\frac{1}{D_1^{a_1} D_2^{a_2}} =
\frac{\Gamma(a_1+a_2)}{\Gamma(a_1)\Gamma(a_2)}
\int_0^1 dx \ 
\frac{ x^{a_1 -1} (1-x)^{a_2 -1} }{ \left[D_1 x+ D_2 (1-x)\right]^{a_1+a_2} }
\end{equation}

\subsection{Integrals}

\begin{equation}
\int \frac{d^{2\omega} l}{(2\pi)^{2\omega}} \ 
\frac{1}{\left(l^2+M^2\right)^A} = 
\frac{1}{(4\pi)^\omega \Gamma(A)} 
\frac{\Gamma(A-\omega)}{\left(M^2\right)^{A-\omega}}
\label{eqn:identity1}
\end{equation}

\begin{equation}
\int \frac{d^{2\omega} l}{(2\pi)^{2\omega}} \ 
 \frac{l_\mu l_\nu}{\left(l^2+M^2\right)^A} = 
\frac{1}{(4\pi)^\omega \Gamma(A)} 
\frac{\delta_{\mu\nu}}{2} 
\frac{\Gamma(A-1-\omega)}{\left(M^2\right)^{A-1-\omega}}
\label{eqn:identity3}
\end{equation}

\subsection{Mathematica Code for $\overset{\rm dim-reg}{K(p)} - \overset{\rm op-reg}{K(p)}$}
This code was composed replacing the operator regularization poles by hand ($1/\varepsilon^2 \to \alpha_2$ and $1/\varepsilon \to \alpha_1$) as this made for faster running; it was also found that the final double integral ran much faster in an older version of Mathematica (version 4), as recent versions are more careful about assumptions.

\begin{verbatim}
arg = 
  p^2 y(1-y) + m^2 ( 1-y + y/(x(1-x)) );

dimreg = 
  -Gamma[2e]/(4Pi)^(4-2e) 1/e *
   (x(1-x))^-e y^e D[(1-y) arg^(-2e), y];
opreg = 
  -1/(4Pi)^4 Gamma[4e]/(Gamma[e+1])^2 1/Gamma[2e+2] 1/(2e) *
   (x(1-x))^-e y^(2e) D[(1-y)^(2e+1) arg^(-4e), y];

dimregexp = 
  Series[ dimreg, {e,0,0}];
opregexp = 
  Series[ opreg, {e,0,0}];

pole2diff = 
  1/e^2 Coefficient[ dimregexp, e,-2] - 
   a2 Coefficient[ opregexp, e,-2];
pole1diff = 
  1/e Coefficient[ dimregexp, e,-1] - 
   a1 Coefficient[ opregexp, e,-1];
finitediff = 
  Coefficient[ dimregexp, e,0] - 
   Coefficient[ opregexp, e,0];

totaldiff = 
  pole2diff + pole1diff + finitediff;

result =
  Integrate[ totaldiff, {x,0,1},{y,0,1}]
\end{verbatim}
If needed (for speed), the problem can be simplified further by first showing the result is independent of $p$ (by showing the derivative with respect to $p^2$ is zero) then setting $p$ to zero before proceeding.

\end{document}